\documentclass[runningheads,a4paper]{llncs}

\usepackage{amssymb}
\usepackage{graphicx}

\usepackage{multirow}

\usepackage{hyperref}

\usepackage{url}
\urldef{\mailsc}\path|{alfred.hofmann, ursula.barth, ingrid.haas, frank.holzwarth,|
\urldef{\mailsb}\path|chuliang.weng@huawei.com, RNambiar@cisco.com|
\urldef{\mailsa}\path|{zhuyuqing, wl, zhanjianfeng, zhangjinchao, chenxingzhen}@ict.ac.cn,|

\begin{document}

\mainmatter  

\title{BigOP: Generating Comprehensive Big Data Workloads as a Benchmarking Framework}

\titlerunning{BigOP}

\author{Yuqing Zhu\thanks{The corresponding author.}%
\and Jianfeng Zhan \and Chuliang Weng$^{\sharp}$ \and Raghunath Nambiar$^{\diamond}$ \and Jinchao Zhang \and Xingzhen Chen \and Lei Wang}
%
\authorrunning{Yuqing Zhu et al.}

\institute{State Key Laboratory of Computer Architecture (Institute of Computing Technology, Chinese Academy of Sciences), $~^{\sharp}$Huawei, $~^{\diamond}$Cisco\\
\{zhuyuqing, zhanjianfeng, zhangjinchao, chenxingzhen, wanglei\_2011\}@ict.ac.cn, $~^{\sharp}$chuliang.weng@huawei.com, $~^{\diamond}$RNambiar@cisco.com\vspace{-12pt}
}

\toctitle{BigOP}
\tocauthor{Yuqing Zhu et al.}
\maketitle

\begin{abstract}
Big Data is considered proprietary asset of companies, organizations, and even nations. Turning big data into real treasure requires the support of big data systems. A variety of commercial and open source products have been unleashed for big data storage and processing. While big data users are facing the choice of which system best suits their needs, big data system developers are facing the question of how to evaluate their systems with regard to general big data processing needs. System benchmarking is the classic way of meeting the above demands. However, existent big data benchmarks either fail to represent the variety of big data processing requirements, or target only one specific platform, e.g. Hadoop.

In this paper, with our industrial partners, we present Big\textbf{\emph{OP}}, an end-to-end system benchmarking framework, featuring the abstraction of representative \textbf{\emph{O}}peration sets, workload \textbf{\emph{P}}atterns, and prescribed tests.  BigOP is part of an open-source big data benchmarking project, \textbf{\emph{BigDataBench}}. BigOP's abstraction model not only guides the development of BigDataBench, but also enables automatic generation of tests with comprehensive workloads.

We illustrate the feasibility of BigOP by implementing an automatic test generation tool and benchmarking against three widely used big data processing systems, i.e. Hadoop, Spark and MySQL Cluster. Three tests targeting three different application scenarios are prescribed. The tests involve relational data, text data and graph data, as well as all operations and workload patterns. We report results following test specifications.\vspace{-6pt}
\end{abstract}
\section{Introduction}\vspace{-6pt}
Companies, organizations and countries are taking big data as their important assets, as the era of big data has inevitably arrived. But drawing insights from big data and turning big data into real treasure demand an in-depth extraction of its values, which heavily relies upon and hence boosts the deployment of massive big data systems.

Big data owners are facing the problem of how to choose the right system for their big data processing requirements, while a variety of commercial and open source products, e.g., NoSQL databases \cite{nosql}, Hadoop MapReduce \cite{hadoop}, Spark \cite{spark}, Impala \cite{impala}, Hive \cite{hive} and Redshift \cite{redshift}, have been unleashed for big data storage and processing. On the other hand, big data system developers are in need of application-perspective evaluation methods for their systems. Benchmarking is the classic way to direct the evaluation and the comparison of systems.

Though many well-established benchmarks exist, e.g. TPC series benchmarks \cite{tpc} and HPL benchmarks \cite{hpl}, no widely accepted benchmark exists for big data systems. Some benchmarks targeting \emph{big data systems} appear in recent years \cite{bigbench,hibench,ycsb,cloudsuite}, but they are either for a specific platform or covering limited workload patterns.

Together with our industrial partners, we present in this paper an end-to-end system benchmarking framework BigOP, which enables automatic generation of tests with comprehensive workloads for big data systems. We build BigOP for our urgent need to benchmark big data systems. BigOP is part of a comprehensive big data benchmarking suite \emph{\textbf{BigDataBench}}\cite{bigdatabench}, which is already used by our collaborators in testing architecture, network and energy efficiency of big data systems. The development of BigDataBench is guided by BigOP's abstraction model.

BigOP features an abstracted set of \textbf{\emph{O}}perations and \textbf{\emph{P}}atterns for big data processing. We work out the abstraction after considering the powerful representativeness of the five primitive relational operators \cite{relational} and the 13 computation patterns summarized in a report by a multidisciplinary group of well-known researchers \cite{berkeleyview}. The operations are extended from the five primitive relational operators, while the workload patterns are summarized based on general big data computation characteristics.

\begin{figure}[!t]
      \centering
      \includegraphics[width=0.6\textwidth]{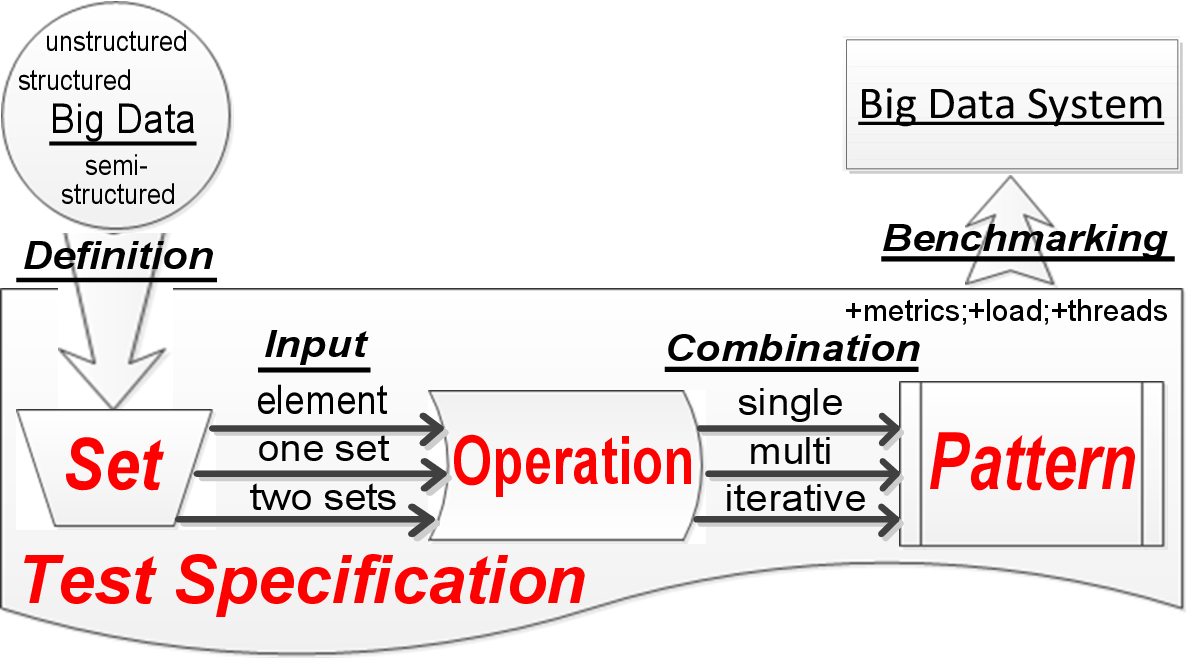}
      \caption{Overview of BigOP.}\vspace{-18pt}
      \label{fig:abs} 
\end{figure}

Figure \ref{fig:abs} demonstrates an overview of BigOP. In BigOP, a benchmarking test is specified as a prescription for one application or a range of applications. A prescription includes a subset of operations and processing patterns, a data set, a workload generation method, and the metrics. The subset of operations and processing patterns are selected from BigOP's whole abstraction set. The data set can be obtained from real applications or generated through widely-obtainable tools. The workload generation method describes how operations are issued from clients, e.g, the number of client threads, the load, etc. The metrics can include the test duration, request latency metrics, and the throughput. With BigOP, a prescribed test can be implemented over different systems for comparison. System users can also prescribe a test targeting their specific applications.\vspace{-12pt}

BigOP leaves the choice of data set to users because the \emph{variety} of big data makes a predefined data set for benchmarking irrelevant. Besides, data sets are usually related to the processing performance in big data scenarios, for example, highly isolated webpages vs. highly linked webpages for PageRank computations. The size of the chosen data set is required to be larger than the total memory size of the system under test (SUT) so that the volume characteristic of big data is covered. The velocity characteristic of big data can be represented in the workload generation specification. That is, BigOP design takes the three properties of Big Data into consideration.

After presenting the design of BigOP (\textbf{Section \ref{sec:BigOP}}), we give three test prescriptions targeting three different application scenarios as an example. We benchmark against three widely used big data processing systems, i.e. MySQL cluster, Hadoop\cite{hadoop}+HBase\cite{hbase} and Spark \cite{spark}, using BigOP (\textbf{Section \ref{sec:eval}}). The tests involve relational data, text data and graph data, as well as all operations and workload patterns. We discuss workload representativeness by comparing YCSB \cite{ycsb}, TPC-DS \cite{tpcds} and BigBench \cite{bigbench} to BigOP. We summarize related work (\textbf{Section \ref{sec:related}}) and conclude (\textbf{Section \ref{sec:conclude}}) in the end.\vspace{-12pt}
\section{BigOP Design}\vspace{-6pt}
\label{sec:BigOP}%
\subsection{Overview}\vspace{-6pt}
BigOP is an end-to-end system benchmarking framework with a big data processing operation and pattern (\textbf{\emph{OP}}) abstraction. Comprehensive workloads can be specified and thus constructed based on the \textbf{\emph{OP}} abstraction.

An adequate level of abstraction and the end-to-end execution model leaves space for various system implementations and optimizations. Benchmarks with these two properties enable comparisons among different kinds of systems servicing the same goals. The success of TPC benchmarks \cite{tpc} demonstrates this fact \cite{benchmarkmodel}. Therefore, BigOP takes an end-to-end benchmarking model. Benchmarking workloads are applied by clients issuing requests through interfaces. Requests are sent through a network to the system under test (SUT). Metrics are measured at the client side.

The success of TPC benchmarks also highlights the importance of benchmarking systems with functions of abstraction and the functional workload model \cite{benchmarkmodel}. Functions of abstraction are basic units of computation occurring frequently in applications,  while the functional workload model includes functions of abstraction, the representative application load, and the data set. Functions of abstraction is the core. To generalize functions of abstraction for BigOP, we abstract a set of operations and patterns common to big data processing.

Furthermore, big data embodies great variety. So do big data applications. Therefore, BigOP allows users to flexibly specify data sets and workloads in test prescriptions. The test prescription allows for application-specific features, as well as comparisons across systems.

\subsection{Operation and Pattern Abstraction for Big Data Processing}
Before going into the details of BigOP's operation and pattern abstraction, we first review some facts about big data processing.

The concept of \emph{set} in relational algebra \cite{relational} is still effective in big data processing scenarios, e.g. the MapReduce \cite{mapreduce} model, which plays an important role in big data processing. In the model, a large piece of data is transformed into a set of elements denoted by key-value pairs in the \emph{map} stage. The \emph{reduce} stage does further processing over the mapped set. We thus adopt the general concept of \emph{set} in BigOP. Elements in a set can be uniquely identified.

Data accesses to memory and disk, as well as across network, must be considered in big data benchmarking. Memory size plays an important role as for system performance. As technology improves, the starting data size of TPC-H increases along with the obtainable amount of memory. Thus, benchmarks must consider the whole system, including the system composition. Besides, big data systems can consist of not only nodes, but also datacenters, due to the huge volume of big data. Thus, communication must also be considered in benchmarks. Furthermore, the huge volume of big data and the resulting processing complexity demand distribution and parallelization of computation tasks. Otherwise, the time required for big data processing would be intolerably long. Hence, BigOP requires the data set size to exceed the total memory size of the SUT, but a single element in the data set must fit into a single node's memory to be processed; or, the element can at least be read serially and transformed into a set of smaller elements for processing.\vspace{-12pt}

\subsubsection{Operation Abstraction}
Considering the above facts, we first abstract big data processing operations into three categories, i.e., \emph{element operation}, \emph{single-set operation} and \emph{double-set operation}. Element operation can be computed based on an individual element, which might require only local memory access. Single-set operations are computed based on elements of one set. Double-set operations require input  from two sets of data. We did not include multi-set operations, which can be composed by combining multiple double-set operations. The more sets are involved in a processing task, the more demands for data accesses involving memory, disk and network communication are there. Operations from the three categories can be combined and permutated to meet complex processing requirements. Table \ref{tbl:op} illustrates the three categories of operations.
\begin{table}[!t]
\centering
\caption{Basic Operations}\vspace{-6pt} %
\label{tbl:op}%
\begin{tabular}{|p{1.9in}|p{2.8in}|}
\hline
\multicolumn{1}{|c|}{\textbf{Categories}} & \multicolumn{1}{|c|}{\textbf{Typical Operations}}\\
\hline
\hline
\textbf{Element Operation} & put, get, delete; transform; filter\\%
\hline
\multirow{2}{*}{\textbf{Single-Set Operation}} & project, order by\\
& aggregation(min,max,sum,median,average)\\
\hline
\textbf{Double-Set Operation}& union, difference, cross product\\
\hline
\end{tabular}\vspace{-12pt}
\end{table}

BigOP adopts the five primitive operators from \emph{relational algebra} \cite{relational}, which also takes a \emph{set}-based perspective. They are filter(select), project, cross product, set union, and set difference. The five primitive operators are fundamental in the sense that omitting any of them causes a loss of expressive power. Many other set operations can be defined in terms of these five. The \emph{filter} operation is an \emph{element} operation because it can operate a set element on given conditions with only element-local information. The \emph{project} operation is in the \emph{single-set} operation category, requiring the set information. The \emph{union}, \emph{difference}, and \emph{cross product} fall in the \emph{double-set} operation category.

The basic data access operations of \emph{put}, \emph{get} and \emph{delete} are in the element operation category. We also add a \emph{transform} operation to this category because it is common to turn a big element into a set of elements or another element, as demonstrated by the Map usages of MapReduce. \emph{transform} is user-defined. Most big data are unstructured data, therefore \emph{transform} is important to define data-specific computations.

We also include the commonly used \emph{order by} and \emph{aggregation} operations in the single-set operation category. \emph{order by} is equal to \emph{sort}, a fundamental database operation noted by Jim Gray \cite{berkeleyview}. Quite a few benchmarks have been built based on \emph{sort} \cite{sort}. \emph{Aggregation} is included because it is widely recognized important operation to turn sets into numerals.\vspace{-12pt}
\begin{table}[!t]
\centering
\caption{Workload Patterns}\vspace{-6pt} %
\label{tbl:pattern}%
\begin{tabular}{|p{2.3in}|p{2.3in}|}
\hline
\multicolumn{1}{|c|}{\textbf{Patterns}} & \multicolumn{1}{|c|}{\textbf{Example Workloads}}\\
\hline
\hline
\textbf{Single-Operation Processing}& any abstracted operation\\
\hline
\textbf{Multi-Operation Processing} & operation combinations, SQL queries\\ 
\hline
\textbf{Iterative Processing} & graph traversal, finite state machines\\
\hline
\end{tabular}\vspace{-12pt}
\end{table}
\subsubsection{Pattern Abstraction}
The abstracted operations can be combined into more complex processing tasks following some patterns. We summarize three patterns as demonstrated with examples in Table \ref{tbl:pattern}. The three patterns are \emph{single-operation}, \emph{multi-operation} and \emph{iterative} processing patterns. The single-operation pattern contains only a single operation in a processing task, while the multi-operation and iterative patterns can have multiple operations in a task. The inclusion of multiple operations allows the big data system to make a whole optimization plan for all operations in the task. The difference between multi-operation processing and iterative processing patterns is whether the exact number of operations to be executed is known beforehand. Iterative processing patterns only provide stopping conditions (which can be specified as user-defined functions), thus the exact number of operations can only be figured out in running time.

Different from SQL queries, the processing patterns in Table \ref{tbl:pattern} can result in more than one set. Furthermore, the element definition of a data set relies on the \emph{transform} operation, instead of \emph{schema}. For example, a text document can be \emph{transformed} into a set with \emph{word} elements or with \emph{sentence} elements. While element operation can be processed locally, global optimization techniques can be employed for single-set and double-set operations, as well as for multi-operation and iterative processing patterns.

The choice of the operations in Table \ref{tbl:op} is in no way complete, but it is representative enough to represent a broad range of processing workloads when used with the patterns of Table \ref{tbl:pattern}. Besides, we think the efforts in benchmarking should be incremental and evolving. That is, more basic operations can be added to Table \ref{tbl:op} in the future, as well as more patterns to Table \ref{tbl:pattern}.\vspace{-12pt}
\begin{table}[!t]
\centering
\caption{Test Prescriptions without Workload Generation Method Specifications}\vspace{-6pt} %
\label{tbl:tests}%
\begin{footnotesize}
\begin{tabular}{|p{0.8in}|p{1.2in}|p{1.5in}|p{1.2in}|}
\hline
    & \multicolumn{1}{|c|}{\textbf{Fast}} & \multicolumn{1}{|c|}{\textbf{Log}}& \multicolumn{1}{|c|}{\textbf{PageRank}}\\
    & \multicolumn{1}{|c|}{\textbf{Storage}} & \multicolumn{1}{|c|}{\textbf{Monitoring}}& \multicolumn{1}{|c|}{\textbf{Computation}}\\
\hline
\hline
\textbf{Operations}& put, get, delete &  put, get, filter, &get, transform, filter,\\
                                     & union & aggregation & order by\\
\hline
\textbf{Patterns}& single-operation & single- and multi-operation &all patterns\\
\hline
\textbf{Data Set}& randomly generated & real server logs & randomly generated\\
                                & structured data         &                                   & directed graph             \\
\hline
\textbf{Metrics} & throughput & request latency statistics& test duration \\
\hline
\end{tabular}\vspace{-12pt}%
\end{footnotesize}
\end{table}
\subsection{Prescriptions and Prescribed Tests}\vspace{-6pt}
Each benchmarking test is specified by a prescription. Thanks to the abstraction of processing operations and patterns, a prescribed test can be implemented over different systems for comparison. A prescription includes a subset of operations and processing patterns, a data set, a workload generation method, and the measured metrics. The subset of operation and processing patterns is selected from BigOP's whole abstraction set. The data set can be taken from real applications or generated through widely-obtainable tools. The size of the chosen data set is required to be larger than the total memory size of the SUT, so that the volume characteristic of big data is covered. The workload generation method describes how operations are issued from clients. The velocity characteristic of big data can be represented in the workload generation specification. The measured metrics can be the duration of the test, request latency statistics, and the throughput.

We instantiate three prescribed test examples in Table \ref{tbl:tests}, from the simplest to the most complex. The first and the third examples\footnote{Referred as Cloud OLTP and PageRank workloads respectively in BigDataBench.} are taken from BigDataBench, while the second example is constructed from a common application scenario. The workload generation methods are not included in the prescription for space consideration. Instead, we describe them respectively in the following, together with an introduction to the application corresponding to each example.\vspace{-6pt}

\begin{example}
\textbf{Fast Storage}. Applications make frequent requests of data storage. This is the most basic scenario of big data acquisition. The data set is generated using BDGS' data generation tool \cite{bdgs}. The workload generation combines YCSB's workloads. The throughput of operations is the key metric.\vspace{-6pt}
\end{example}

\begin{example}
\textbf{Log Monitoring}. An application monitors its services by logs. Applications like user and server activity monitoring can be represented by this prescription. The workload generation is specified as follows. \emph{put} is applied to some random log entries at a speed of 5000 operations per second (ops) till the total data size exceeds twice of the total system memory. Simultaneously, \emph{get+filter+aggregation} is applied to the data set based on some random filtering condition for a \emph{sum} result continuously.\vspace{-6pt}
\end{example}

\begin{example}
\label{exmpl:pagerank}
\textbf{PageRank Computation}. This is a core computation in the widely deployed Internet service. It is also a representative computation of graph applications. In workload generation, \emph{get+filter+ transform} is applied to the data set \emph{iteratively} till a given \emph{condition} is met. The \emph{order by} operation, a.k.a. \emph{sort}, is executed to get the final result.\vspace{-6pt}
\end{example}

In constructing test prescriptions, we can adjust factors in the prescription according to our needs. For example, to test for data scalability, we can increase the workload of \emph{put}. Similarly, to test more complicated computations, we can increase the number of combinations for \emph{multi-operation} and \emph{iterative} processing patterns, as well as defining a sophisticated \emph{transform} function.\vspace{-6pt}
\begin{figure}[!t]
      \centering
      \includegraphics[width=0.6\textwidth]{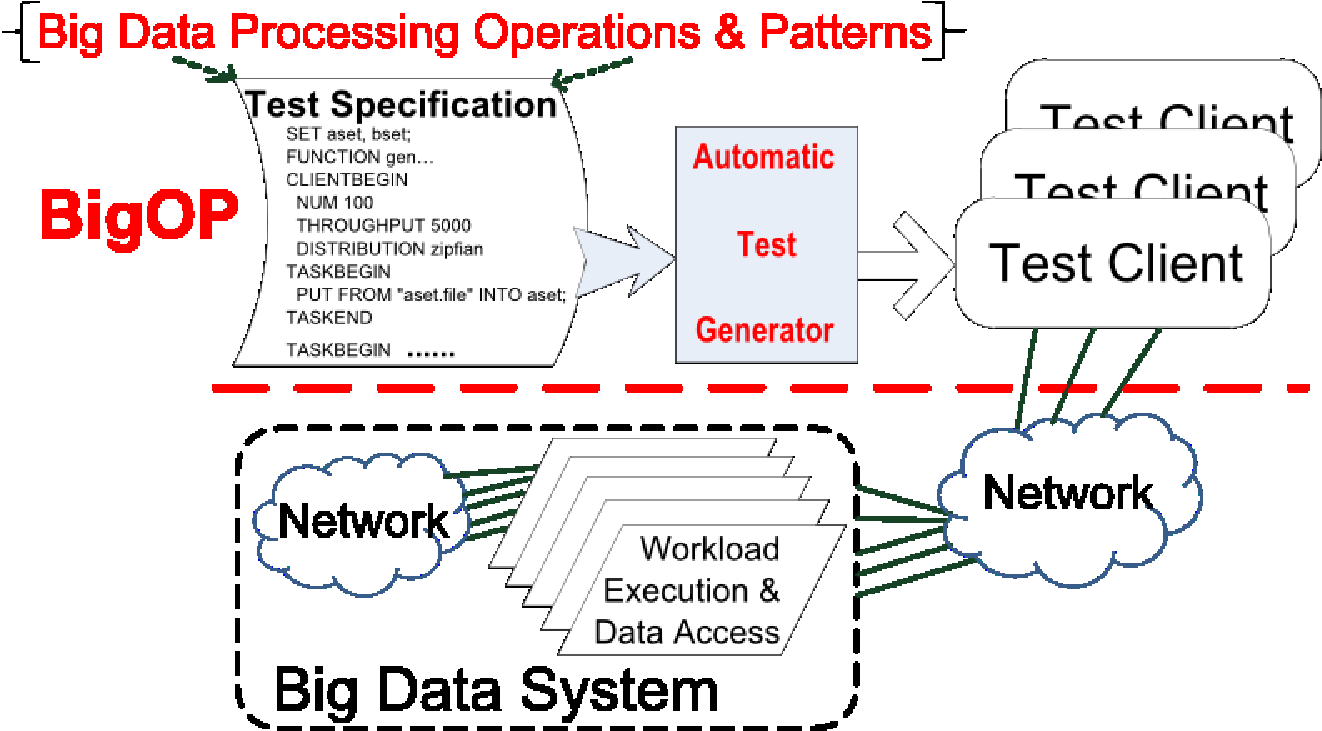}
      \caption{Testing process and functional components of BigOP.}
      \label{fig:test} 
\end{figure}%
\section{Evaluation}\vspace{-6pt}%
\label{sec:eval}%
Based on the BigOP framework, we implement an automatic test generator, which can generate tests from prescriptions. Figure \ref{fig:test} demonstrates the testing procedure of the evaluation.

We benchmark three widely used big data processing systems, i.e. MySQL cluster (SUT1), Hadoop+HBase (SUT2) and HDFS+Spark (SUT3). All the three SUTs are deployed over five physical nodes connected with 1Gbps network. Each node has 32 GB memory and a processor with six 2.40GHz cores. The operating system is Centos release 5.5 with Linux kernel 2.6.34. Among our three examples, we only choose the second and the third for the evaluation, since the first example has been extensively tested in other benchmarks.
\setlength{\textfloatsep}{0.6\baselineskip}
\begin{figure}[!t]
      \centering
      \includegraphics[width=0.8\textwidth]{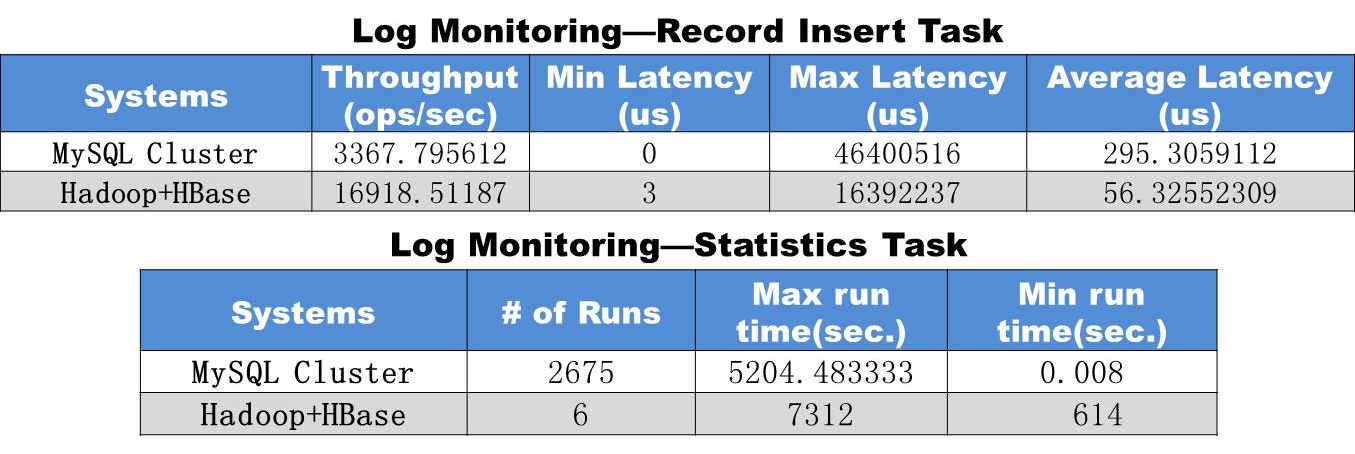}
      \caption{System performances under log monitoring workloads. (The zero min latency of SUT1 is due to the batch commit mode.)}
      \label{fig:log} 
\end{figure}%

\textbf{Log Monitoring}. Figure \ref{fig:log} demonstrates the resulting performances of SUT1 and SUT2. SUT2 excels under the frequent record insert task as expected. SUT1 performs much better in statistics computation tasks because of the long starting time of jobs in SUT2. For complex multi-operation tasks, SUT1 is very likely to excel the others as well. The reason is as follows. Even though MapReduce-like systems including Hadoop, Shark and Hive support complex user defined functions, they have not considered optimizations on multi-operation and iterative processing patterns. On the contrary, traditional databases are strictly SQL compliant and heavily optimized for relational queries, which usually contain single-operation and multi-operation processing patterns.

\textbf{Pagerank Computation}. We generate 0.5 million pages and 3.7 million links using an open-source tool \cite{bdgs}, resulting in more than 250GB data (including page contents), which is larger than the total system memory. The running time of the task over SUT2 is \textbf{\emph{363 seconds}}, while that over SUT3 is \emph{\textbf{96 seconds}}. We also run this test over SUT1 through a stored procedure, which contains costly large-scale joins leading to intolerable test durations. The indication here is \emph{twofold}. First, distributed in-memory computation is effective for the \emph{iterative} computation pattern, while frequent disk accesses and network communications can be costly. Second, there is still much optimization space for distributed computation in relational database, especially when the iterative pattern is involved.

We report a small fraction of evaluation results here due to the page limit. Further benchmarking results can be found on the BigDataBench webpage\footnote{http://prof.ict.ac.cn/BigDataBench/}.\vspace{-12pt}

\subsubsection{Discussion}

Existent big data benchmarks only cover part of BigOP's abstraction of processing operations and patterns. We take YCSB \cite{ycsb}, TPC-DS \cite{tpcds}, and BigBench \cite{bigbench} for example. YCSB is mainly for NoSQL database benchmarking. Its workload consists of only \emph{put} and \emph{get} operations, though which can be combined by the \emph{multi-operation} pattern to form \emph{scan} and \emph{read-modify-write} operations. TPC-DS covers all abstracted operations and the first two patterns, except for the \emph{iterative} pattern. It targets a single application domain, i.e., decision support applications. The involved data is structured data. Thus, it is not as flexible in suiting different benchmarking requirements as BigOP. BigBench extends TPC-DS. It adds new data types of semi-structured data and unstructured data, but it still does not include the \emph{iterative} pattern.\vspace{-6pt}

\section{Related Work}\vspace{-9pt}%
\label{sec:related}%
BigBench \cite{bigbench} is the recent effort towards designing a general big data benchmark. BigBench focuses on big data analytics, thus adopting TPC-DS as the basis and adding atop new data types like semi-/un-structured data, as well as non-relational workloads like sentiment queries. Although BigBench has a complete coverage of data types, it targets only a specific big data application scenario, not covering the variety of big data processing workloads.

The AMPLab of UC Berkeley also proposes a big data benchmark \cite{ampbig} in recent years. It is the systems of Spark \cite{spark} and Shark \cite{shark} that inspire the design of the benchmark, which thus targets real-time analytic applications. The benchmark not only has a limited coverage of workloads, but also covers only relational data.

Industrial players also try to develop their benchmark suites. Yahoo! release their cloud  benchmark specially for data storage systems, i.e. YCSB\cite{ycsb}. Having its root in cloud computing, YCSB is mainly for scenarios like that in the \emph{Fast Storage} example. The characteristics of and the diverse application workloads for big data are not considered in YCSB. CALDA\cite{calda} effort represents a micro benchmark for Big Data analytics. It has compared Hadoop-based analytics to a row-based RDBMS one and a column-based RDBMS one.

There exist also benchmarks targeting a specific platform, e.g. Hadoop \cite{hadoop}. HiBench \cite{hibench} is a widely-known benchmark for Hadoop MapReduce. Hence, its four categories of workloads are limited to MapReduce based processing. It exploits stochastic methods to generate data for its workloads. However, its randomly generated data misses various features of real big data. Besides, its choice of workloads lacks of coverage, as well as solidly founded grounds. Gridmix \cite{gridmix} and Pigmix \cite{pigmix} are also two benchmarks specially designed for Hadoop MapReduce. They include a mix of workloads, including sort, grep and wordcount. They are also suffering from incomplete coverage of data and workloads.

Across different research fields, there are multiple famous and well established benchmarks. The TeraSort or GraySort Benchmark \cite{sort} considers the performance and the cost involved in sorting a large number of 100-byte records. The workload of this benchmark is too simple to cover the various needs of big data processing. SPEC \cite{spec} works well over standalone servers with a homogeneous architecture, but is not suitable for the emerging big data platforms with large-scale distributed and heterogeneous components. TPC \cite{tpc} series of benchmarks are widely accepted for database testing, but only consider structured data. PARSEC \cite{parsec} is a well-known benchmark for shared-memory computers, thus not suited for big data applications that mainly take a shared-nothing architecture.\vspace{-6pt}

\section{Conclusion}\vspace{-9pt}%
\label{sec:conclude}%
BigOP is targeted at big data systems that support part of or all of its processing operation and pattern abstractions. Users of BigOP can flexibly construct tests through prescriptions for their application-specific or general benchmarking needs. Benchmarking tests can be constructed based on BigOP's abstracted operations and patterns. Big data systems that can implement the same prescribed test are comparable. System users can prescribe tests targeting at their specific application scenarios, while system developers can carry out general tests with all abstracted operations and patterns mixed and randomly generated in the workload. We design BigOP as a benchmarking framework in considering the variety of big data and its applications, which we believe is a good trade-off between benchmarking flexibility and conformity to real big data processing requirements.\vspace{-6pt}

\section*{Acknowledgments}
This work is supported in part by the State Key Development Program for Basic Research of China (Grant No. 2014CB340402) and the National Natural Science Foundation of China (Grant No. 61303054).


%
\bibliographystyle{abbrv}
{\footnotesize
\bibliography{ref}

\begin{thebibliography}{10}

\bibitem{redshift}
Amazon redshift service.
\newblock http://aws.amazon.com/cn/redshift/.

\bibitem{ampbig}
Amplab big data benchmark.
\newblock https://amplab.cs.berkeley.edu/benchmark/.

\bibitem{impala}
Cloudera impala.
\newblock
  http://www.cloudera.com/content/cloudera/en/products-and-services/cdh/impala.html.

\bibitem{gridmix}
Gridmix.
\newblock http://hadoop.apache.org/mapreduce/docs/current/gridmix.html.

\bibitem{hadoop}
Hadoop project.
\newblock http://hadoop.apache.org/.

\bibitem{hbase}
Hbase project.
\newblock http://hbase.apache.org/.

\bibitem{hive}
Hive project.
\newblock http://hive.apache.org/.

\bibitem{hpl}
Hpl benchmark home page.
\newblock http://www.netlib.org/benchmark/hpl/.

\bibitem{nosql}
Nosql databases.
\newblock http://nosql-database.org/.

\bibitem{pigmix}
Pigmix.
\newblock https://cwiki.apache.org/confluence/display/PIG/PigMix.

\bibitem{sort}
Sort benchmark home page.
\newblock http://sortbenchmark.org/.

\bibitem{spec}
Standard performance evaluation corporation (spec).
\newblock http://www.spec.org/.

\bibitem{tpc}
Transaction processing performance council (tpc).
\newblock http://www.tpc.org/.

\bibitem{berkeleyview}
K.~Asanovic, R.~Bodik, B.~C. Catanzaro, J.~J. Gebis, P.~Husbands, K.~Keutzer,
  D.~A. Patterson, W.~L. Plishker, J.~Shalf, S.~W. Williams, et~al.
\newblock The landscape of parallel computing research: A view from berkeley.
\newblock Technical report, UCB/EECS-2006-183, EECS Department, University of
  California, Berkeley, 2006.

\bibitem{parsec}
C.~Bienia, S.~Kumar, J.~Singh, and K.~Li.
\newblock The parsec benchmark suite: Characterization and architectural
  implications.
\newblock In {\em Proc. of PACT 2008}, pages 72--81. ACM, 2008.

\bibitem{benchmarkmodel}
Y.~Chen, F.~Raab, and R.~H. Katz.
\newblock From tpc-c to big data benchmarks: A functional workload model.
\newblock Technical Report UCB/EECS-2012-174, EECS Department, University of
  California, Berkeley, Jul 2012.

\bibitem{relational}
E.~F. Codd.
\newblock A relational model of data for large shared data banks.
\newblock In {\em Pioneers and Their Contributions to Software Engineering},
  pages 61--98. Springer, 2001.

\bibitem{ycsb}
B.~F. Cooper, A.~Silberstein, E.~Tam, R.~Ramakrishnan, and R.~Sears.
\newblock Benchmarking cloud serving systems with ycsb.
\newblock In {\em Proc. of SoCC 2010}.

\bibitem{mapreduce}
J.~Dean and S.~Ghemawat.
\newblock Mapreduce: simplified data processing on large clusters.
\newblock {\em Communications of the ACM}, 51(1):107--113, 2008.

\bibitem{shark}
C.~Engle, A.~Lupher, R.~Xin, M.~Zaharia, M.~J. Franklin, S.~Shenker, and
  I.~Stoica.
\newblock Shark: fast data analysis using coarse-grained distributed memory.
\newblock In {\em Proc. of SIGMOD 2012}, pages 689--692. ACM, 2012.

\bibitem{cloudsuite}
M.~Ferdman, A.~Adileh, O.~Kocberber, S.~Volos, M.~Alisafaee, D.~Jevdjic,
  C.~Kaynak, A.~Popescu, A.~Ailamaki, and B.~Falsafi.
\newblock Clearing the clouds: A study of emerging workloads on modern
  hardware.
\newblock {\em Architectural Support for Programming Languages and Operating
  Systems}, 2012.

\bibitem{bigbench}
A.~Ghazal, M.~Hu, T.~Rabl, F.~Raab, M.~Poess, A.~Crolotte, and H.-A. Jacobsen.
\newblock Bigbench: Towards an industry standard benchmark for big data
  analytics.
\newblock In {\em Proc. of SIGMOD 2013}. ACM, 2013.

\bibitem{hibench}
S.~Huang, J.~Huang, J.~Dai, T.~Xie, and B.~Huang.
\newblock The hibench benchmark suite: Characterization of the mapreduce-based
  data analysis.
\newblock In {\em Proc. of ICDEW 2010}, pages 41--51. IEEE, 2010.

\bibitem{bdgs}
Z.~Ming, C.~Luo, W.~Gao, R.~Han, Q.~Yang, L.~Wang, and J.~Zhan.
\newblock Bdgs: A scalable big data generator suite in big data benchmarking.
\newblock {\em arXiv:1401.5465}, 2014.

\bibitem{calda}
A.~Pavlo, E.~Paulson, A.~Rasin, D.~J. Abadi, D.~J. DeWitt, S.~Madden, and
  M.~Stonebraker.
\newblock A comparison of approaches to large-scale data analysis.
\newblock In {\em Proc. of SIGMOD 2009}, pages 165--178, New York, NY, USA,
  2009. ACM.

\bibitem{tpcds}
M.~Poess, R.~O. Nambiar, and D.~Walrath.
\newblock Why you should run tpc-ds: a workload analysis.
\newblock In {\em Proc. of VLDB 2007}, pages 1138--1149. VLDB Endowment, 2007.

\bibitem{bigdatabench}
L.~Wang, J.~Zhan, C.~Luo, Y.~Zhu, Q.~Yang, Y.~He, W.~Gao, Z.~Jia, Y.~Shi,
  S.~Zhang, C.~Zhen, G.~Lu, K.~Zhan, and B.~Qiu.
\newblock Bigdatabench: a big data benchmark suite from internet services.
\newblock Accepted by HPCA 2014.

\bibitem{spark}
M.~Zaharia, M.~Chowdhury, T.~Das, A.~Dave, J.~Ma, M.~McCauley, M.~Franklin,
  S.~Shenker, and I.~Stoica.
\newblock Resilient distributed datasets: A fault-tolerant abstraction for
  in-memory cluster computing.
\newblock In {\em Proc. of NSDI 2012}, pages 2--2. USENIX Association, 2012.

\end{thebibliography}
} %

\end{document}